# Head and Eye Control in Persons with Low Vision during Urban Navigation

Mahya Beheshti 1,2*, JR Rizzo 1-6*, Sarah Bergquist-Kosumi 7, Ajayrangan Kasturirangan 7, Sharvari Deshpande 7 , Todd Hudson 1,5,6

*co-first authors

1)      Department of Rehabilitation Medicine, NYU Langone Health, New York, USA

2)      Department of Mechanical & Aerospace Eng., NYU Tandon School of Engineering, New York, USA

3)      Department of Ophthalmology, NYU Langone Health, New York, USA

4)      Department of Neurology, NYU Langone Health, New York, USA

5)      Department of Biomedical Engineering, NYU Tandon School of Engineering, New York, USA

6)      Institute for Excellence in Health Equity, New York University Grossman School of Medicine, New York, USA

7)      Department of Computer Science and Engineering, NYU Tandon School of Engineering, Brooklyn, NY 11201, USA

*co-first authors

Correspondent Author:

Mahya Beheshti

Beheshti.mahya@gmail.com

# Abstract

Low vision involves a range of visual impairments that significantly impact daily activities, particularly navigation in urban environments. Individuals with low vision often develop adaptive strategies to compensate for visual deficits, relying on head movements to bring objects into their remaining functional field of vision. Research suggests that they focus on road surface markings and building edges to aid in wayfinding and collision avoidance. However, urban navigation presents additional challenges, as obstacles, moving hazards, and tripping dangers may enter their visual loss field, increasing the risk of injury. Traditional eye movement studies are typically conducted in controlled laboratory settings with fixed head positions, limiting the understanding of head-eye coordination in real-world environments. To bridge this gap, we designed a naturalistic, "free-head" experiment using eye-tracking technology to examine head and eye movement patterns during urban navigation. Participants with low vision were compared to a control cohort without visual impairment to test the hypothesis that eye and head movements become decoupled in visually impaired individuals.

Findings indicate that individuals with peripheral field loss exhibit significant eye-head decoupling, while those with acuity loss demonstrate more synchronized movements. Results for individuals with central field loss were inconclusive but revealed distinct movement patterns. These insights provide valuable direction for rehabilitation strategies, assistive- mobility technologies, and urban design improvements. By expanding research on eye-head coordination, this study contributes to the development of interventions that enhance safety, mobility, and independence for individuals with low vision in complex urban environments.

Keywords

Visual Impairment, Eye-Tracking, Eye-Head Coupling, Sensory-motor integration

**Introduction:**

Navigating urban environments is a fundamental yet challenging task for daily life . The control of line of sight by the central nervous system, which encompasses both eye and head movements, plays a crucial role in perceiving, localizing, and recognizing objects in space. A key component of visuomotor behavior during navigation is eye-head coupling, which governs the synchronization between eye and head movements. Eye-head coupling refers to the coordination between these movements, enabling individuals to stabilize their gaze and orient efficiently within their surroundings. In persons without sight loss, eye-head coupling stabilizes vision and enables rapid gaze shifts. Typically, a gaze shift begins with a saccadic eye movement, followed by a coordinated head movement to optimize target acquisition (Freedman, 2008). This coupling minimizes excessive eye movements, balancing accuracy and speed (Salvador & Chan, 2007).

Low vision encompasses a spectrum of visual impairments, including central and peripheral vision loss, as well as reduced acuity, each affecting navigation differently. Research suggests that individuals with visual impairments rely on adaptive strategies such as head movements and environmental cues, including road markings and building edges, to compensate for reduced visual input (Lappi, 2015; Owsley, McGwin, Lee, Wasserman, & Searcey, 2009; Salvador & Chan, 2007).However, the extent to which these strategies vary based on the type and severity of visual impairment remains largely unexamined, particularly in real-world settings. For example, those with peripheral vision loss often increase reliance on head movements, whereas those with central vision loss may exhibit erratic eye movement patterns due to restricted foveal vision (Lappi, 2015).

Given the unpredictability of urban environments—characterized by moving objects, architectural structures, and variable lighting conditions—understanding how individuals with low vision adapt their visuomotor strategies is essential for improving mobility and safety (Leigh & Zee, 2015). Previous research on eye and head coordination in low vision has been limited, with a primary focus on eye movements and less consideration for head movements, leading to a gap in understanding how head movements contribute to visual navigation. Previous studies conducted in controlled laboratory environments were limited and devoid of human visual behavior assessments in real-world scenarios. Conducting research in naturalistic environments is essential to determine the actual visual strategies employed by individuals when confronted with complex real-world tasks (Lappi, 2015). Limited research exists on the coupling mechanisms between eye and head movements in individuals with low vision, particularly regarding their utilization of the oculomotor system, sensory inputs, and motor controls to perceive and navigate their surroundings effectively.

Advancements in eye-tracking technology offer new opportunities to study eye-head coupling in real-world contexts. Portable eye-tracking systems allow for precise analysis of gaze behavior, saccadic patterns, and head movement dynamics in urban environments (Chen, Liu, Kojima, Huang, & Arai, 2021). While research suggests that individuals with low vision employ distinct scanning strategies compared to sighted individuals, the extent to which eye-head coupling differs across various types of visual impairment remains underexplored (Tachiquin et al., 2021).

The objective of this study was to investigate how different types of visual impairment influence eye-head coupling during urban navigation. By leveraging free-head eye-tracking methodologies, we test the hypothesis that decoupling is present to variable degrees in visual loss and linked to visual phenotype.

**Methodology**

Data Engineering

The pupil labs eye tracker provides data regarding eye movements, head movements, and the raw video during the experiment. Combining the eye and head movement data is important to analyze any metric involving the two measures. A data engineering pipeline had been designed to fetch the pupil labs data from the database and process it into actionable formats for further analysis. This resulted in two datasets, one having information for each timestamp and another summarized for each saccade (Figure 1). The data engineering pipeline involves the following:

- Timestamp Level Data

  - Merge raw data of eye and head movement according to matching global timestamps.
  - Classify each timestamp according to the task being performed during the experiment.
  - Derive metrics like orientation and velocity for each timestamp from existing variables.

- Saccade Level Data

  - Filter data based on data classified as fixations and a velocity threshold.
  - Assign a unique identifier for each saccade.
  - Calculate metrics for each saccade including peak velocity, duration etc.

The minimum viable product (MVP) is the result of integrating the eye and head data for 1 subject. The purpose of using the MVP is to gain familiarity with the data by conducting exploratory data analysis. Using only 1 subject accelerates the process by focusing on a smaller, more manageable data subset.

Upon satisfactory completion of exploratory data analysis, the natural succession in the methodology is to scale the study to analyze data involving all subjects. Processed datasets of the head and eye movement were obtained by aggregating the minimum viable product of all subjects as well as including information about the subject including the age and cohort.

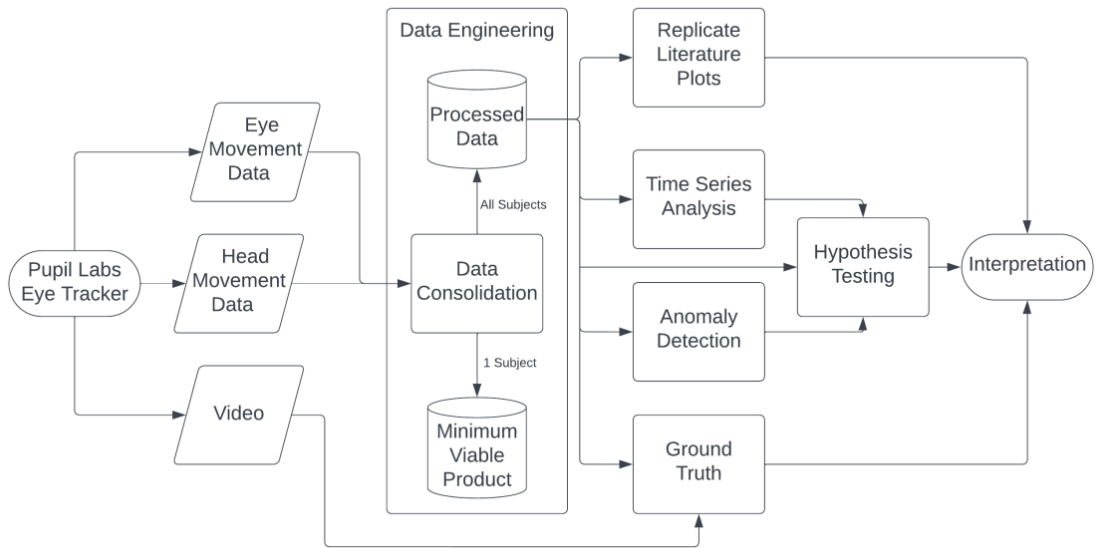

*Figure 1: Methodology Flowchart*

Data Collection:

Fifty-one subjects participated in this study, 31 healthy control and 21 individuals with visual impairments. (See Table 1 for Counts per Vision Type and Task). After signing consent subjects walked to the subway wearing eye tracking glasses.

Tasks were recorded using the Pupil Invisible mobile eye-tracking headset (Pupil Labs). The pupil labs eye tracker provides data regarding eye movements, head movements (using the embedded Inertial Measurement Unit (IMU)) and the raw video during the experiment. The nanosecond unit eye movement data includes hand labeled 'task' information (walking downstairs/upstairs, across the street, etc.), IMU head movement readings (position, pitch, etc.), and eye gaze readings (position, elevation, azimuth)

|  | **Healthy Control** | **Acuity** | **Central** | **Peripheral** | **Total** |
|---|---|---|---|---|---|
| Number of Subjects | 30 | 5 | 9 | 7 | 51 |
| Task |  |  |  |  | Total |
| Crossing street | 778 | 237 | 302 | 263 | 1580 |
| Downstairs | 322 | 114 | 102 | 253 | 791 |
| In subway station | 23 | 38 | 34 | 101 | 196 |
| Indoor | 1022 | 153 | 115 | 137 | 1427 |
| No Task | 15374 | 46 | 88 | 117 | 15625 |
| Obstacle | 2305 | 566 | 744 | 683 | 4298 |
| Sidewalk no scaffolding | 7773 | 2309 | 2872 | 2676 | 15630 |
| Sidewalk with scaffolding | 655 | 188 | 245 | 399 | 1487 |
| Upstairs | 108 | 54 | 81 | 115 | 358 |
| Waiting for stop sign | 593 | 239 | 278 | 343 | 1453 |
| Total Saccades | 28953 | 3944 | 4861 | 5087 | 42845 |

Table 1: Participant counts and task-level saccades by vision type and navigation condition.

Analytic Approach

Replicate Standard Plots: Previous research provided a solid framework of standard basis for patterns of eye and head movements, e.g., the main sequence relationship. A subset of charts and metric approaches were applied to this analysis per cohort, providing a way to compare differences between each low vision cohort and the healthy control.

Time Series Analyses: Temporal pattern comparison of the eye and head was the key analytic approach needed to identify the degree of head and eye movement coupling/decoupling. Our approach to temporal pattern comparison relied on several metrics:

• Dynamic Time Warping (DTW): In contrast to other measures of similarity or distance between signals, DTW was the preferred time series analysis approach because it can effectively compare and align movements, even in cases where there are nonlinearities in the relationship between time-series. The output of DTW is a distance/similarity metric, which was used to compare the overall match between eye and head time-series for each low vision cohort and healthy control (low values indicate better coupling whereas higher values indicate decoupling).

• Cross-Correlation: To evaluate the temporal distributions of head and eye velocity, the maximum, lag and overall similarity (area) are calculated. These summaries explain the time delay offsets and provide insights into the average temporal coupling or decoupling of the head compared to the eye.

• Power Spectral Density (PSD): Quantifies the power (energy per unit time) contained at different frequencies contained within the time-series, the PSD provides an understanding of the repetitive behavior of head and eye movements. The PSD can reveal whether repetitive behaviors are consistent (power concentrated within a small range of frequencies) or variable (power diffused across a wide range of frequencies). The total power, represented by the area under the curve in the PSD, is an essential metric in interpreting the results. It signifies the total energy contained in the signal across all frequencies. Higher total power indicates that the time series contains more overall activity.

Anomaly Detection: To better understand which cohort's eye and head movements fall outside expected patterns, this study applied a density-based spatial clustering of applications with noise (DBSCAN) to identify global outliers, as well as Local Outlier Factor (LOF) to understand outliers based on neighborhood density. These results of anomaly detection were used to find the proportion of each cohort that is anomalous, as well as a filtering mechanism to reduce noise in the final saccade data set.

Hypothesis Testing: Statistical evidence is required to establish a difference between multiple cohorts in the experiment. Therefore, this study used a Mann-Whitney U test, a non-parametric method, instead of a standard students T-test because the assumptions of normality and homogeneity cannot be made. Many metrics were calculated on eye, head, eye+head positional and velocity movements. Each cohort was compared to the healthy control and key results were identified at the 95% confidence level ($p<0.05$).

Ground Truth Verification The participant overlay videos were examined to ensure the accuracy of the results, particularly in instances where the outcomes were unexpected. It helped in shaping the recommendations as it offered insights into the experiences of each participant while performing specific tasks. Figure 1 illustrates the complex dynamics of eye and head movements during the task downstairs, given their particular type of vision loss. This explains the differential patterns of each individual subject, thereby providing a more comprehensive understanding of how vision loss influences their navigation strategy during tasks such as going downstairs.

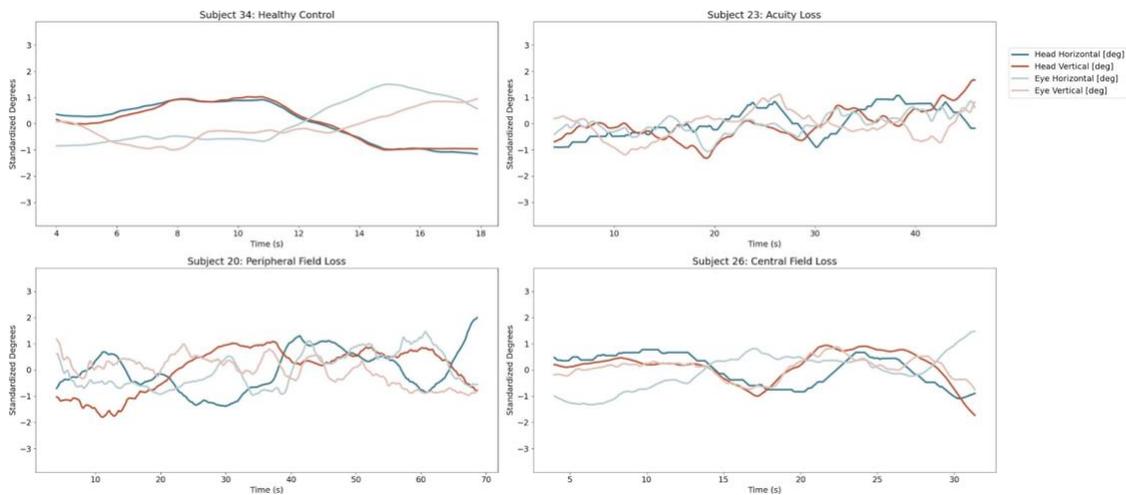

*Figure 2:Position Plot for Task Downstairs at Subject Level*

Analytic Limitations

The limitations of this study include the lack of quantifying the severity of vision loss among participants, which could influence the results. Due to constraints in existing technology, the use of IMU for tracking head position inevitably leads to error accumulation over time (Kothari et al., 2020). The IMU data from the Pupil Labs eye tracker introduced drift over time in gyroscope readings. The sample size was skewed, with fewer subjects in the low vision compared to healthy control groups, and many of the cases were not verified. Not all tasks were included for some subjects, which could limit the applicability of comprehensive comparisons. The absence of fixed targets in a naturalistic experiment, as opposed to laboratory settings, makes it challenging to determine the subjects' focus of gaze, and constituting an additional source of noise.

Risk Mitigation

To mitigate potential risks, careful attention was dedicated into ensuring data privacy, best practices for this field of study, and data cleanliness. This included confirming the sponsor's adherence to appropriate disclosures, securing a data release agreement, and verifying the absence of personally identifiable information within the data. A thorough literature review informed the project's foundation and helped focus on gaps in the research space. Regular sponsor check-ins yielded invaluable iterative feedback from experts. The MVP and exploratory data analysis stages served to validate the approach by addressing issues like IMU drift and unexpected data noise. Robust quality assurance measures, such as the ground truth verification minimized the risk of a flawed statistical analysis.

**Results:**

*Overview of Findings*

Replication of the main sequence relationship confirmed the validity of the saccade detection across all cohorts. Analyses revealed phenotype-specific differences in eye and head dynamics. Dynamic Time Warping (DTW) was used as the primary outcome measure to quantify overall eye–head synchronization. (Table 2)

*Eye-Head Coupling (Primary Outcome)*

DTW quantified the alignment between eye and head velocity and position traces. The density map for eye-head movements is illustrated in Figure 5. Higher DTW distances indicated greater decoupling, while lower distances indicated tighter coupling. Individuals with peripheral vision loss exhibited significantly larger DTW distances compared to controls, demonstrating decoupling of eye and head movements. Participants with reduced acuity showed smaller DTW distances, reflecting stronger eye–head coupling. Central vision loss participants displayed variable patterns, with no significant group-level difference, suggesting heterogeneity within this cohort.

*Eye Movement Metrics*

Eye movement analyses revealed further phenotype-specific differences. Peripheral loss participants had amplitudes similar to controls but compressed in range, while central loss participants exhibited wider amplitudes, suggesting increased effort to achieve target acquisition. Peripheral loss participants demonstrated slower, longer saccades at larger amplitudes, whereas central loss participants exhibited quicker saccades at narrower amplitudes. Acuity loss participants showed longer saccades for the same amplitude. All cohorts followed the main sequence, but peripheral loss showed faster peak velocities than controls, while acuity loss produced slower peak velocities.

*Head Movement Metrics*

Head movement analyses revealed complementary adaptations. Peripheral loss participants demonstrated a wider vertical range of head movement and higher power spectral density (PSD), consistent with more intense vertical scanning. Central loss participants exhibited narrower vertical ranges but higher PSD, indicating localized but effortful head adjustments. Acuity loss participants demonstrated less variable head movements with overall lower PSD. Both peripheral and acuity loss cohorts showed slower peak head velocities compared to controls, suggesting that head adjustments were not able to match the altered demands of the saccadic behavior.

*Power Spectral Density (PSD)*

PSD analyses highlighted the intensity and frequency of both eye and head movements. Peripheral loss cohorts displayed higher PSD values, reflecting more frequent and robust scanning activity. Central loss cohorts showed narrower ranges of movement but with high PSD, consistent with constrained yet effortful adjustments. Acuity loss cohorts demonstrated the lowest PSD values, consistent with less intense movement overall.

| Cohort | Eye Movements | Head Movements | PSD Findings |
| --- | --- | --- | --- |
| Peripheral Vision Loss | Slightly less horizontal amplitude range, saccades often faster at larger amplitudes | Wider vertical head movement range, horizontal head movements less intense; slower head peak velocity | Higher PSD overall, especially vertical → indicates more frequent/intense scanning |
| Central Vision Loss | Wider amplitude range, distinctive vertical trajectories, less stereotyped | Narrower vertical head movement range, less variable | High PSD despite narrower range → suggests localized |

|  | horizontal movements | head movements overall | effortful adjustments |
| --- | --- | --- | --- |
| Acuity Loss | Smaller, more stereotyped set of horizontal movements, longer saccades at same amplitude | Head movements primarily above horizon, less variable horizontal/vertical adjustments, slower velocities | Lowest PSD overall → indicates less intense head and eye movement |

Table 2: Summary of phenotype-specific eye movements, head movements, and PSD findings across vision loss cohorts.

*Integration Across Measures*

These findings demonstrate that altered saccadic and head movement properties translate into distinct compensatory strategies. Peripheral vision loss leads to an eye-dominant strategy, characterized by rapid scanning and eye–head decoupling. Reduced acuity results in stronger coupling of eyes and head, consistent with a stabilization strategy to maximize detail. Central vision loss produces variable adaptations that reflect heterogeneity in strategies for managing foveal deficits. DTW provides the unifying methodology that links these eye-only and head-only findings into an interpretation of phenotype-specific compensatory strategies.

Amplitude

The total degree the participants moved during a saccade differ from the healthy control for all of the cohorts. The peripheral loss cohort had amplitudes that were most similar to persons without sight loss, but overall range maximum range was lower by 15 degrees. Those with central loss showed slightly wider amplitudes, suggesting more movement is required to focus (Figure 3).

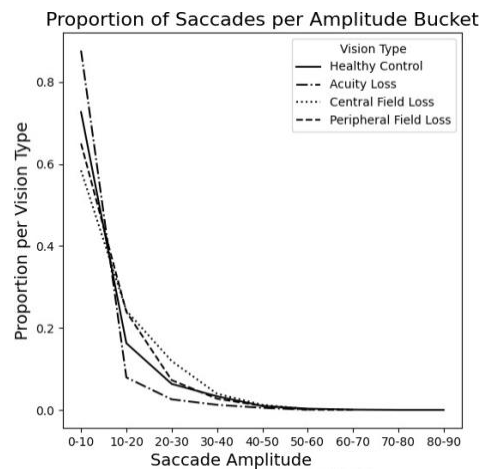

*Figure 3: Frequency of saccade amplitudes in a naturalistic setting*

Duration

The peak velocity and duration of saccades as a function of the amplitude of the saccade are displayed in Figure 4. A clear difference in duration, peak velocity, and the duration of the peak velocity was observed for peripheral vision loss in comparison to controls. The saccades are slower and take longer to complete, particularly as the amplitude increases. Subjects with central vision loss exhibit quicker saccades at narrower amplitudes, and slightly longer durations at wider amplitudes. Subjects with acuity loss tend to display longer saccades than controls for the same amplitude.

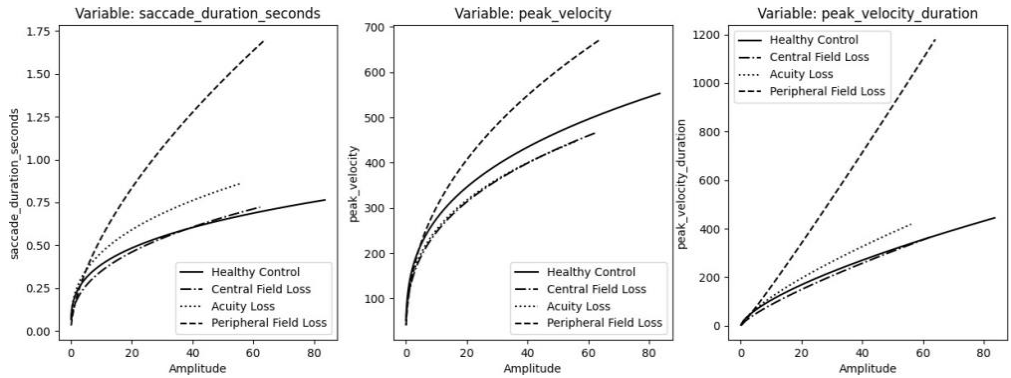

Figure 4: Peak Velocity, Duration vs Amplitude per Vision Type

Position

The density map for eye-head movements is illustrated in Figure 5. Gaze is directed down in the Peripheral Vision Loss cohort proportionally more often than in healthy control and exhibited a wider eye elevation range, along with a higher PSD, indicating more intense and frequent movement. The Central Vision Loss cohort demonstrated an even wider range and a high PSD. The Acuity Loss cohort had their movement primarily above the horizon with the statistical measures confined within a small positive range. Their PSD was the lowest among the cohorts suggesting less overall movement intensity. The central vision loss cohort exhibited a significantly smaller average range. The acuity loss cohort also displayed a smaller average range, indicative of a more stereotyped set of horizontal movement amplitudes. Both the central and acuity loss cohorts demonstrated less intense eye movements, as indicated by lower PSD. In the peripheral vision loss cohort, a slightly wider range of vertical head movement is observed. PSD is higher indicating more overall head movement. Central vision loss cohort has a narrower vertical head movement range, suggesting a less variable head movement pattern. The high PSD values suggest robust and active head movements. For horizontal head movements, individuals with peripheral and central vision loss exhibit less intense movements with lower PSD, whereas those with acuity loss show notably more intense head movements.

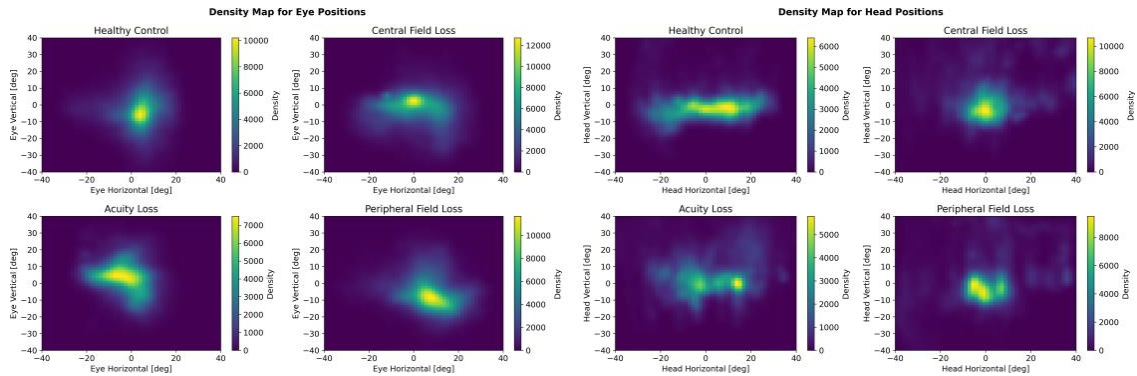

Figure 5: Density Map for Eye and Head Positions

Velocity

All cohorts' saccade patterns follow the main sequence relationship. However, peak velocities for the Peripheral Loss cohort are faster than the control group, especially as the amplitudes increase. While the Acuity Loss group has peak velocities much slower than the control. Peak velocity of the head does not follow the same pattern for the Peripheral Loss cohort, the head moves on average slower, which could mean it is more difficult to adjust the head to put objects into foveal focus. The Acuity Loss cohort has much lower peak head velocities as well. Overall, when comparing the saccade time series of eye velocity to head velocity, DTW shows decoupling in Peripheral loss, where the eye and the head are moving at disjointed, uncoordinated velocities. Acuity loss DTW shows more eye and head velocity coupling, where the eye and the head are moving in a more coordinated speed, which could be due to the overall narrower amplitudes of the cohort. The Central Vision Loss cohort results were not significant.

Eye and Head Coordination

To measure the overall coordination of the eye and head, DTW time series analysis was applied to two vectors (1 for eye and another for head) containing position (vertical & horizontal) and velocity. The distance output found that the Peripheral Loss cohort was more decoupled compared to the healthy control group, meaning the eye and head positional and velocity movements are more uncoordinated. The opposite is true of the Acuity Loss cohort, where the positional and velocity movements are more coupled, meaning the eye and head are more in sync. Unfortunately, the Central Loss cohort had inconclusive results (not significant at the 99% confidence level).

**Discussion:**

This study examined the coordination of eye and head movements during naturalistic urban navigation in individuals with low vision, using Dynamic Time Warping (DTW) as the primary analytic tool. DTW directly quantified the synchronization of eye and head movement traces, with higher values indicating decoupling and lower values indicating tighter coupling. By focusing on DTW, we can demonstrate that visuomotor strategies differ across different visual phenotypes and these represent specific compensatory adaptations.

For individuals with peripheral vision loss, DTW distances were significantly larger than those of controls, indicating a decoupling of eye and head movements. This group also exhibited longer, faster saccades, while head movements lagged and were slower horizontally. These patterns suggest that peripheral loss prompts individuals to rely more heavily on rapid eye scanning rather than coordinated eye–head adjustments. Prior research showed consistent results indicating that peripheral field restrictions demand expanded exploratory strategies (Lappi, 2015; Barhorst-Cates et al., 2019. In contrast, individuals with reduced acuity demonstrated smaller DTW distances and tighter eye–head coupling. This was accompanied by smaller-amplitude, slower saccades and more synchronized head movements. Their visuomotor strategy featured numerous small-amplitude saccades with lower peak velocities, indicative of frequent incremental adjustments. This coordinated movement likely represents an adaptive response to optimize visual clarity, consistent with previously documented scanning behavior in acuity-impaired populations (Salvador & Chan, 2007).

Central vision loss produced distinctive vertical eye-movement patterns but variable DTW results. Although definitive conclusions regarding eye-head decoupling could not be established for this group, the identified pattern underscores important adaptive adjustments, possibly driven by their central foveal impairment. These novel insights highlight the complexity inherent in adapting visual strategies to the demands of real-world navigation, extending the results of prior laboratory-based research into naturalistic urban contexts(Chen et al., 2021; Tachiquin et al., 2021)

The DTW analysis is particularly important because it integrates the eye- and head-specific findings. The differences in saccade amplitude, velocity and duration demonstrate phenotype specific eye movement adaptations, while power spectral density (PSD) analyses captures head-movement intensity. DTW ultimately shows how these components combine into overall coordination. Thus, DTW provides the clearest evidence that peripheral loss drives decoupling, acuity loss strengthens coupling, and central loss produces variable compensations.

The findings of this study carry critical implications for rehabilitation and urban infrastructure design. The pronounced decoupling observed in peripheral vision loss underscores the necessity for urban environments to reduce the demand for extensive horizontal scanning through careful infrastructure planning. For individuals with acuity loss, urban spaces could better support their incremental visual exploration through clearly distinguishable environmental markers and integrated assistive technologies. Additionally, the unique eye movement patterns identified among individuals with central vision loss emphasize the need for targeted rehabilitation approaches and specialized assistive technologies to support practical visual orientation.

For instance, transportation systems like the New York Metropolitan Transit Authority (MTA) have implemented tactile wayfinding strips on station floors to guide individuals with visual impairments toward exits and transfer points. These additions made by the MTA, combined with high-contrast floor markings and clear edge delineations, can have a huge impact in reducing the burden of horizontal scanning in complex environments. For individuals with reduced acuity, who rely on tightly coupled eye head movements and incremental scanning, urban spaces can be optimized by adding high contrast environmental markers that enhance object recognition and localization. A few examples include boldly colored curb edges, high-contrast crosswalks, and brightly lit signage with large print. These modifications help support incremental visual exploration.

This study extends laboratory based work into real world navigation environments and shows that eye-head coordination is not only altered in low vision, but is also phenotype-specific. DTW provides a measure that links eye-only and head-only findings, strengthening the finding that individuals adopt distinctive visuomotor strategies to navigate complex environments.

**Limitations**

Despite its valuable contributions, this study has several limitations. First, it did not quantitatively assess the severity of visual impairment among participants, which may have influenced observed visuomotor behaviors. This limitation suggests that future studies should consider the severity of visual impairment as a potential influencing factor. Additionally, the use of IMUs introduced drift errors in head movement data, inevitably affecting the accuracy of measurements (Kothari et al., 2020). This limitation underscores the need for improved measurement techniques in future studies. Furthermore, a relatively smaller and uneven sample size among vision-impaired cohorts compared to healthy controls limits the generalizability of the findings. This limitation highlights the need for larger and more balanced sample sizes in future research. Lastly, the inherent lack of fixed visual targets in naturalistic urban scenarios posed challenges for accurately determining gaze focus, introducing additional variability into the data. This limitation suggests that future studies should consider the challenges of naturalistic urban scenarios when designing research methodologies.

Addressing these limitations will be critical in future research. Subsequent studies should quantify impairment severity to explore its impact systematically on eye-head coordination. Investigating how visuomotor strategies evolve following specific rehabilitation interventions across different task complexities and environmental contexts could further enhance intervention effectiveness. Qualitative methodologies like the "think-aloud" technique could yield deeper insights into cognitive demands and perceived difficulties during navigation tasks. Additionally, applying principles from Fitts' Law to assess pedestrian signal detection could enhance the understanding of perceptual and decision-making processes in individuals with low vision.

Continued development and rigorous evaluation of advanced assistive technologies, including wearable devices with GPS, computer vision-based object detection, and real-time semantic visual mapping, represent promising future research and development directions. Policymakers and urban planners should proactively integrate these assistive innovations into public infrastructure to ensure comprehensive accessibility and inclusion.

**Conclusion**

The current research underscores critical differences in eye-head coordination strategies among individuals with peripheral, acuity, and central vision loss, validating the hypothesis that peripheral loss is linked to significant decoupling, while acuity loss exhibits enhanced coupling. These insights are instrumental in informing targeted rehabilitation, assistive technology advancements, and inclusive urban infrastructure design. Continued interdisciplinary research, integrating computational models and rehabilitation strategies, will refine our understanding of the environmental factors influencing navigation in visually impaired populations, ultimately promoting safer, more accessible urban environments.

This study demonstrates that eye–head coordination during urban navigation is not only altered in low vision but also phenotype-specific. Using Dynamic Time Warping as a primary outcome, we show that peripheral vision loss produces significant decoupling of eye and head movements, whereas reduced acuity is characterized by enhanced coupling, and central vision loss yields variable but distinctive adaptations. These results show that altered eye and head dynamics manifest as compensatory strategies tailored to the type of visual impairment. The findings have direct implications for rehabilitation, where training may target phenotype-specific strategies, and for the design of assistive technologies and urban environments that better support safe and efficient navigation. Continued interdisciplinary research, integrating computational models and rehabilitation strategies, will refine our understanding of the environmental factors influencing navigation in visually impaired populations, ultimately promoting safer, more accessible urban environments.

*Statements and Declarations:*

NYU, John-Ross Rizzo, and Todd E. Hudson have financial interests in related intellectual property. NYU owns a patent licensed to Tactile Navigation Tools. NYU, John-Ross Rizzo, and Todd E. Hudson are equity holders and advisors of said company.

*Acknowledgments:*

This research was supported by the National Science Foundation under Grant Nos. ECCS-1928614, CNS-1952180, ITE-2236097, ITE-2345139, and DUE-2129076; by the National Eye Institute and Fogarty International Center under Grant Nos. R21EY033689, R33EY033689, and R01EY036667; by the U.S. Department of Defense under Grant No. VR200130; and by the Institute for Excellence in Health Equity. The content is solely the responsibility of the authors and does not necessarily represent the official views of the National Institutes of Health, National Science Foundation, Department of Defense, or the Institute for Excellence in Health Equity.

**References:**

Chen, Z., Liu, X., Kojima, M., Huang, Q., & Arai, T. (2021). A wearable navigation device for visually impaired people based on the real-time semantic visual SLAM system. Sensors, 21(4), 1536.

Freedman, E. G. (2008). Coordination of the eyes and head during visual orienting. Experimental brain research, 190, 369-387.

Kothari, R., Yang, Z., Kanan, C., Bailey, R., Pelz, J. B., & Diaz, G. J. (2020). Gaze-in-wild: A dataset for studying eye and head coordination in everyday activities. Scientific reports, 10(1), 2539.

Lappi, O. (2015). Eye tracking in the wild: the good, the bad and the ugly. Journal of Eye Movement Research, 8(5).

Leigh, R. J., & Zee, D. S. (2015). The saccadic system. The neurology of eye movements, 169-288.

Owsley, C., McGwin, G., Lee, P. P., Wasserman, N., & Searcey, K. (2009). Characteristics of low-vision rehabilitation services in the United States. Archives of Ophthalmology, 127(5), 681-689.

Salvador, S., & Chan, P. (2007). Toward accurate dynamic time warping in linear time and space. Intelligent Data Analysis, 11(5), 561-580.

Tachiquin, R., Velázquez, R., Del-Valle-Soto, C., Gutiérrez, C. A., Carrasco, M., De Fazio, R., . . . Vidal-Verdú, F. (2021). Wearable urban mobility assistive device for visually impaired pedestrians using a smartphone and a tactile-foot interface. Sensors, 21(16), 5274.